\documentclass[aps,prb,twocolumn,groupedaddress,showpacs]{revtex4}

\usepackage{textcomp,amsmath,amssymb,graphicx,float}
\begin{document}


\title{Superconductivity at $T_{\mathrm{c}} \sim 14$ K in Single Crystalline FeTe$_{0.61}$Se$_{0.39}$}


\author{T. Taen$^{1}$}
\author{Y. Tsuchiya$^{1}$}
\author{Y. Nakajima$^{1, 2}$}
\author{T. Tamegai$^{1, 2}$}
\email[Present address:]{tamegai@ap.t.u-tokyo.ac.jp}
\affiliation{$^{1}$Department of Applied Physics, The University of Tokyo, 7-3-1 Hongo, Bunkyo-ku, Tokyo 113-8656, Japan \\
$^{2}$JST, Transformative Research-Project on Iron Pnictides (TRIP), 7-3-1 Hongo, Bunkyo-ku, Tokyo 113-8656, Japan}


\date{\today}

\begin{abstract}
Single crystalline FeTe$_{0.61}$Se$_{0.39}$ with a sharp superconducting transition at  $T_{\textrm{c}} \sim$ 14 K is synthesized via slow furnace cooling followed by low-temperature annealing. The effect of annealing on the chemical and superconducting inhomogeneities is carefully characterized. We also report resistivity, magnetization, and magneto-optical images of this crystal. Based on the Bean model, critical current density is estimated to exceed $1 \times 10^5$ A/cm$^2$ below 5 K under zero field. Weak fish-tail effect is identified at lower temperatures.
\end{abstract}


\maketitle


Since the discovery of iron-based oxypnictide superconductor
LaFeAsO$_{1-x}$F$_{x}$ with $T_{\textrm{c}} \sim 26$ K\cite{jacs},
the transition temperature has been raised up to 55 K in $R$FeAsO$_{1-x}$F$_x$ ($R$ = rare earths)\cite{cpl1}.
In addition to these iron-oxypnictides, oxygen-free iron pnictides
such as Ba$_{1-x}$K$_x$Fe$_2$As$_2$ ($T_{\textrm{c}} \sim$ 38 K)\cite{prl1} or Li$_{1-x}$FeAs($T_{\textrm{c}} \sim$ 18 K)\cite{sols} were found.
All these iron-pnictides have layered structures with conducting FeAs layers sandwiched between blocking layers.
These discoveries of iron-pnictide superconductors were followed by another new family of iron-based superconductors,
iron-chalcogenides Fe$_{1+\delta}$Se with $T_{\textrm{c}} \sim$ 8 K\cite{pnas}
and its substituted compounds Fe$_{1+y}$(Te$_{x}$Se$_{1-x}$)\cite{prb5}.
In contrast to iron-arsenides, this system has only conducting FeX (X: chalcogens) layers,
equivalent to FeAs layers in iron-arsenides.
Due to its simpler structure compared to iron-pnictides and less toxics nature,
iron-chalcogenides could be more suitable for practical applications despite their lower $T_{\textrm{c}}$.
However, synthesis of high-quality iron-chalcogenides is not straightforward.
Although the first paper reported a clean superconducting phase only in Se deficient sample~\cite{pnas},
detailed studies on polycrystalline samples revealed that excess irons exist and formula should be written as Fe$_{1+\delta}$Se~\cite{prb4}.
In addition, superconductivity is very sensitive to synthesis condition, and only samples with small $\delta$ show superconductivity.
To elucidate superconducting mechanism, high-quality single crystals with reasonable value of critical current density, $J_{\textrm{c}}$, are indispensable.
Such crystals are also important to evaluate whether iron-chalcogenides are suitable for applications utilizing high-current-carrying capability, such as cables and magnets.
Single crystal growth of FeSe has been attempted using NaCl flux,
resulting in tiny crystals with $T_{\textrm{c}} \sim 10.4$ K~\cite{sst1}.
Recently, however, large single crystals of Fe$_{1+y}$Te or Fe$_{1+y}$Te$_{x}$Se$_{1-x}$ with the tetragonal structure have been successfully grown by Bridgman method~\cite{prb3, prb1}.

In this paper, we report synthesis and characterization of high-quality single crystals of FeTe$_{0.61}$Se$_{0.39}$ grown by slow cooling and low-temperature annealing.
The homogeneity of the obtained crystal is examined by compositional analysis and magneto-optical (MO) imaging, in addition to standard characterizations of structure, magnetization, and resistivity.

Single crystalline samples with nominal composition FeTe$_{0.5}$Se$_{0.5}$ were prepared from
Fe grains (Kojundo Chemical Laboratory, 99.99 \%), Te powder (same as above, 99.999 \%),
and Se grains (same as above, 99.999 \%).
More than 10 g of stoichiometric quantities were loaded into a small silica tube with $d_{1} \sim 10$ mm$\phi$, evacuated, and sealed.
This tube was sealed into a second evacuated tube with $d_{2} \sim 20$ mm$\phi$.
The whole assembly was heated up to 1070 \textdegree C for 36 hours,
followed by slow cooling down to 710 \textdegree C at a rate of 6 \textdegree C/h.
After that, it was furnace-cooled to room temperature.
The obtained crystals (as-grown) does not show superconductivity except for a small part as discussed below.
As-grown crystals were then sealed into a small silica ampoule under vacuum and annealed at 400 \textdegree C
for more than 10 days, followed by quenching.
The obtained crystal cleaves perpendicular to $c$ axis.
Typical dimensions of the crystal with flat surface are $5 \times 5 \times 1$ mm$^3$.

The phase purity and the chemical composition of the crystals were confirmed
by powder x-ray diffraction (Cu K$\alpha$, M18XHF, MAC Science) 
and EDX (S-4300, Hitachi High-Technologies equipped with EMAX x-act, HORIBA), respectively.
Magnetization was measured by a commercial SQUID magnetometer (MPMS-XL5, Quantum Design).
The resistivity measurements were performed by four-contact method.
In order to decrease the contact resistance,
we sputtered gold on the contact pads just after the cleavage,
and 25 $\mu$m gold lead wires were attached with silver paste (Dupont 4922).
Magneto-optical images were obtained by using the local field-dependent Faraday effect
in the in-plane magnetized garnet indicator film employing a differential method\cite{nat1,prb2}.

\begin{figure}[tb]
\begin{center}
\includegraphics[width=7cm,clip]{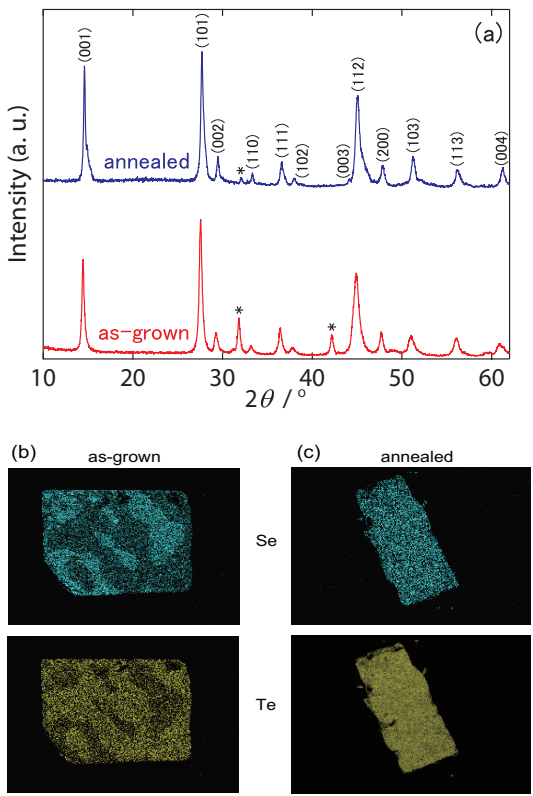}
\end{center}
\caption{(Color online) (a) Powder x-ray diffraction pattern for the as-grown and annealed FeTe$_{0.61}$Se$_{0.39}$ samples. Te/Se distribution probed by EDX analyses in the (b)  as-grown and in (c) annealed samples.}
\label{f1}
\end{figure}

Figure \ref{f1}(a) shows powder x-ray diffraction pattern for the as-grown and annealed samples.
The sample is phase-pure except for a small amount of Fe$_7$Se$_8$.
The amount of Fe$_7$Se$_8$ is strongly reduced after annealing,
which indicates that annealing makes them more homogeneous.
In order to confirm this assumption, we obtained compositional data of several regions
in both samples by EDX analyses.
Figure \ref{f1}(b) and (c) shows the composition map of the as-grown and annealed samples, respectively.
It is clear that inhomogeneous Se/Te distribution is greatly improved by annealing.
The lattice constants determined by x-ray diffraction data are
$a = 3.813$\;\AA, $c = 6.100$\;\AA\; and
$a = 3.798$\;\AA, $c = 6.052$\;\AA\;
for as-grown and annealed samples, respectively.
The latter values of lattice parameters are very close to those reported for Fe$_{1.01}$Te$_{0.54}$Se$_{0.56}$ single crystal~\cite{prb1}.
The chemical composition of the annealed sample determined by EDX analyses is FeTe$_{0.61}$Se$_{0.39}$.

\begin{figure}[tb]
\begin{center}
\includegraphics[width=7cm,clip]{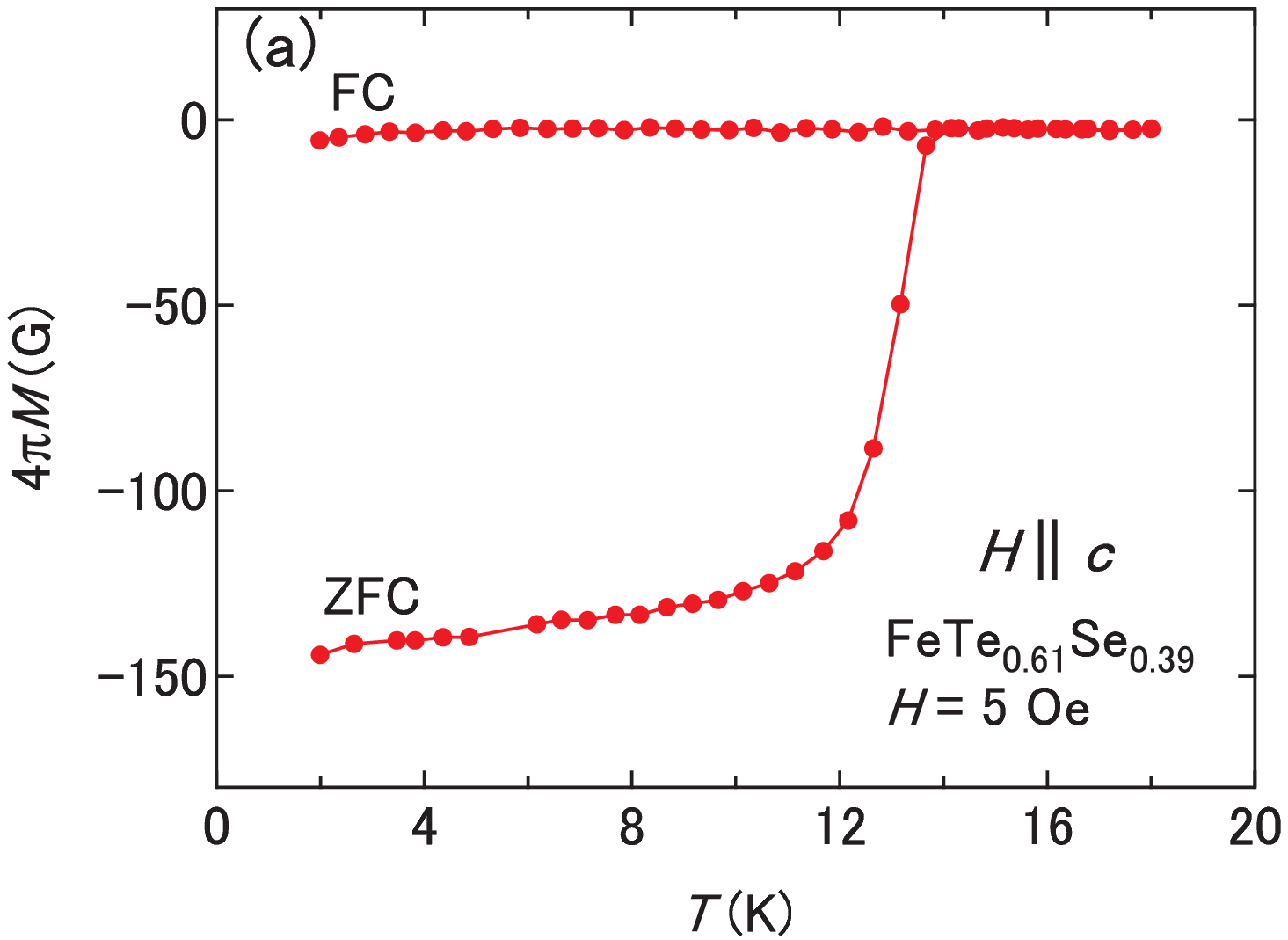}
\includegraphics[width=7cm,clip]{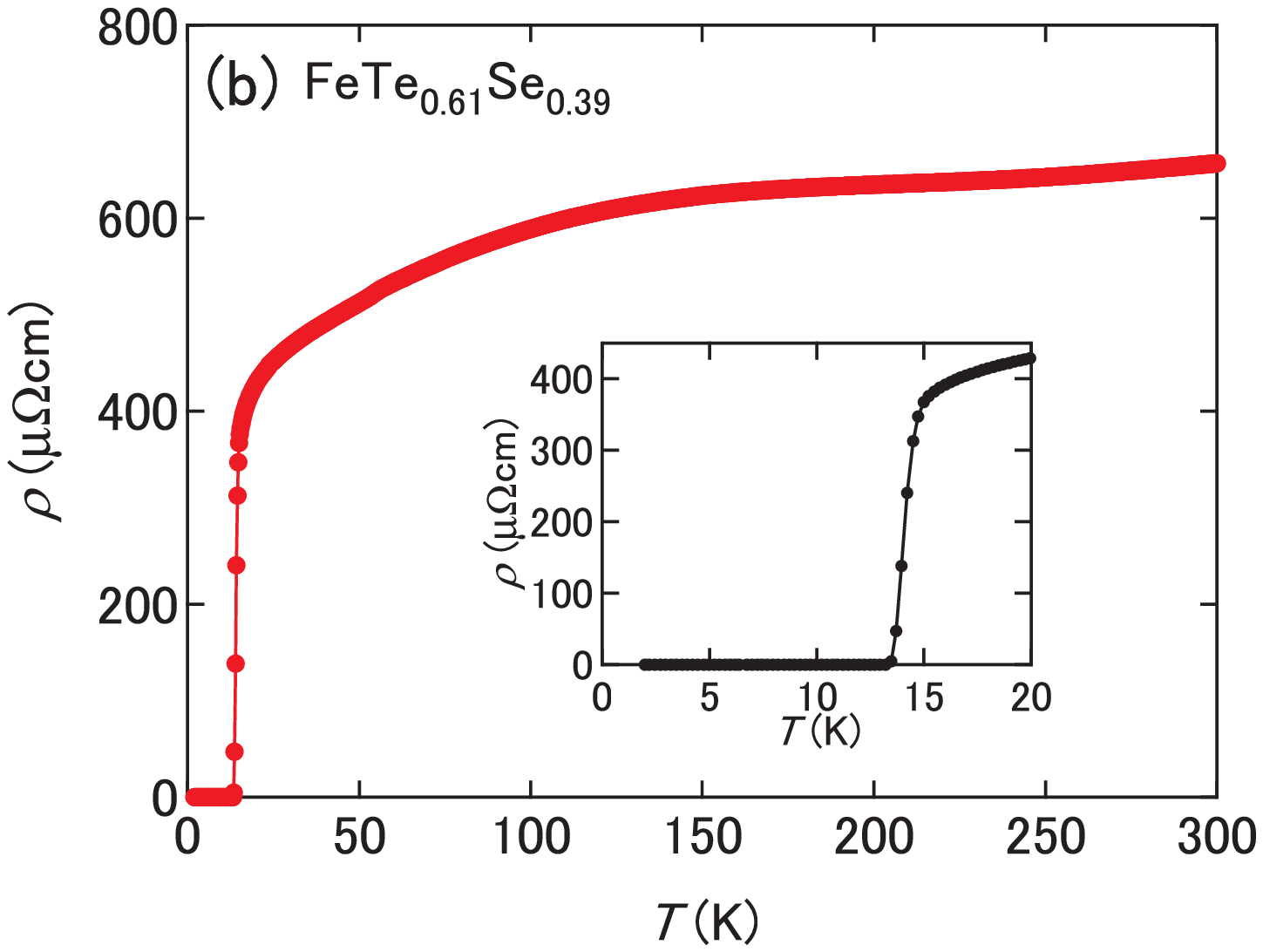}
\end{center}
\caption{(Color online) (a) Temperature dependence of the zero-field-cooled (ZFC) and field-cooled (FC) magnetization at $H$ = 5 Oe in FeTe$_{0.61}$Se$_{0.39}$. (b) Temperature dependence of in-plane resistivity in FeTe$_{0.61}$Se$_{0.39}$. Inset: low-temperature part of the resistive transition.}
\label{f2}
\end{figure}

Figure \ref{f2}(a) shows temperature dependence of zero-field-cooled (ZFC) and field-cooled (FC) magnetization at 5 Oe.
Very sharp transition starting at $T_{\textrm{c}} \sim$ 14 K is observed.
We note that many of the as-grown samples do not show superconductivity.
Figure \ref{f2}(b) shows temperature dependence of in-plane resistivity in FeTe$_{0.61}$Se$_{0.39}$.
Resistivity shows a typical metallic temperature dependence only below 150 K.
In Ref. [10], a similar metallic temperature dependence of resistivity is reported only for crystals with compositions close to FeTe$_{0.5}$Se$_{0.5}$.
In our FeTe$_{0.61}$Se$_{0.39}$, resistivity is nearly constant at higher temperatures.
Such a weak temperature dependence of resistivity is not for a typical metal.
In addition, the value of resistivity at $T_{\textrm{c}}$ is $\sim 400$\; $\mu \Omega$cm,
which is even higher than that in Ba(Fe$_{1-x}$Co$_{x}$)$_2$As$_2$~\cite{jpsj} with disorder in FeAs planes.
This value is twice as large as that for Fe$_{0.99}$Te$_{0.51}$Se$_{0.49}$ reported in Ref. [10].
However, the value of resistivity is very sensitive to the Se content, and temperature dependence of resistivity becomes almost flat or even semiconducting down to $T_{\textrm{c}}$ in crystals with $x > 0.55$~\cite{prb1}.
It is possible that microscopic inhomogeneities still present in our crystal are responsible for such an anomalous temperature dependence and large residual resistivity.

\begin{figure}[tb]
\begin{center}
\includegraphics[width=7cm,clip]{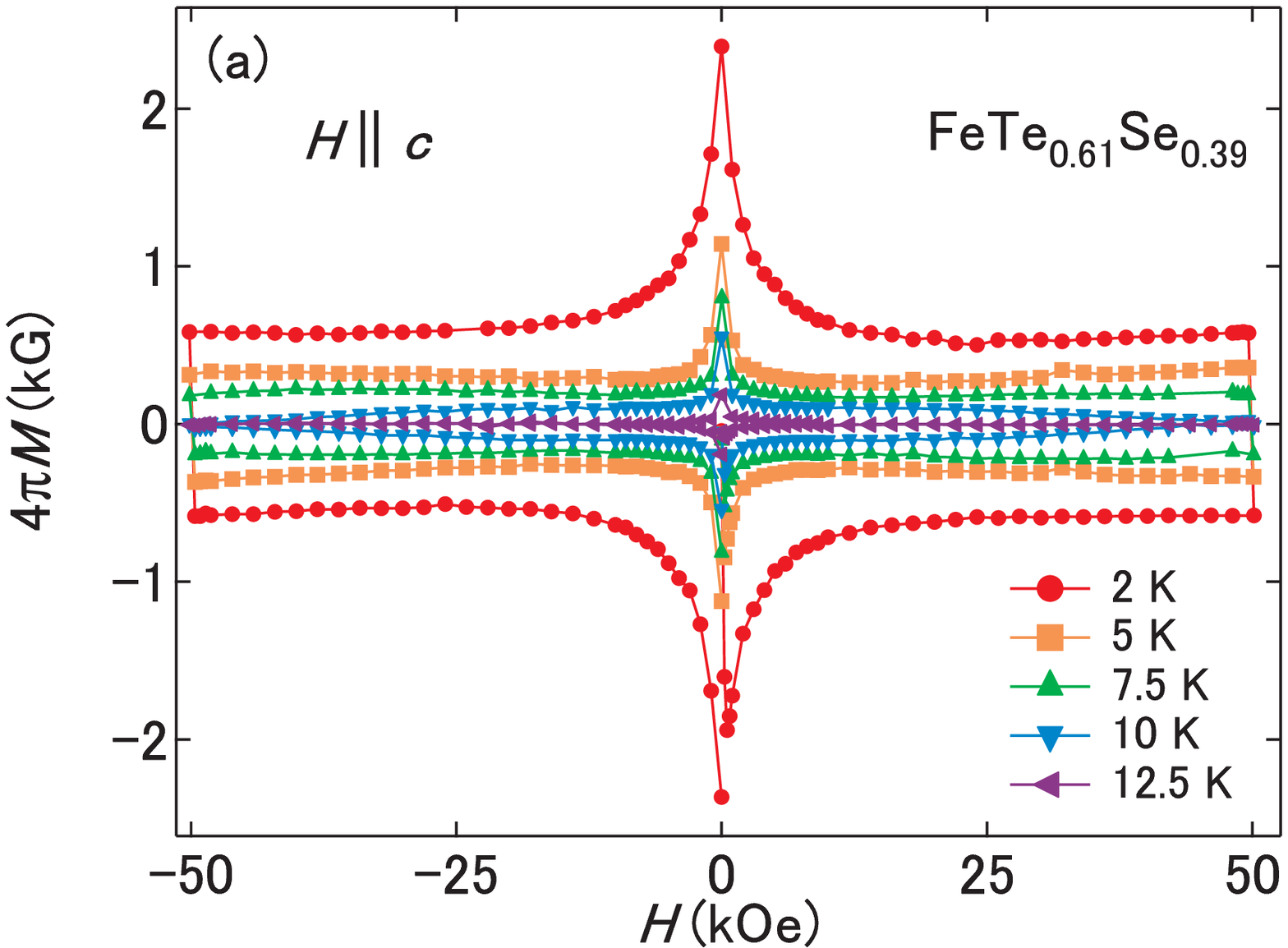}

\includegraphics[width=7cm,clip]{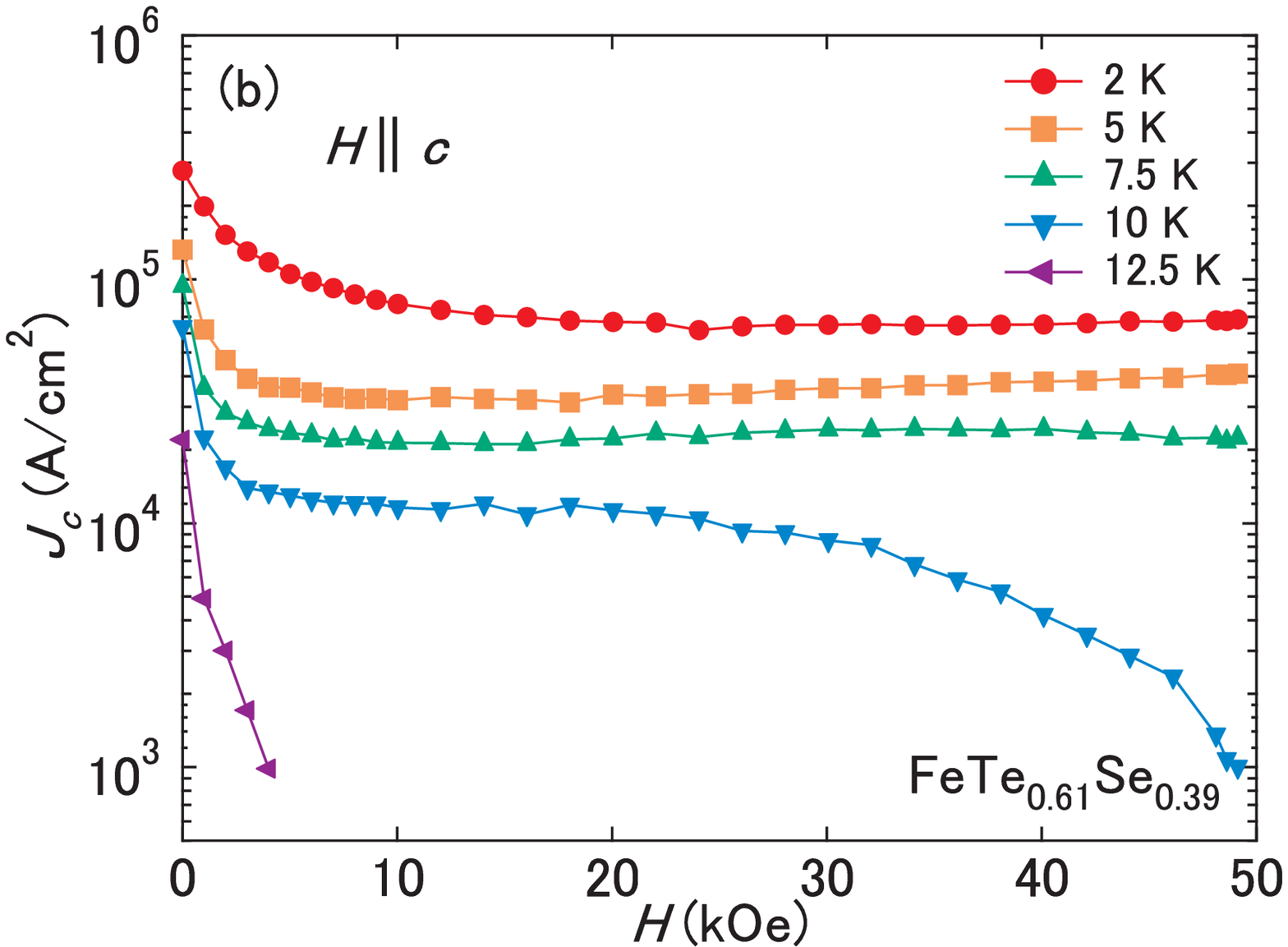}
\end{center}
\caption{(Color online) (a) Magnetic field dependence of magnetization in FeTe$_{0.61}$Se$_{0.39}$ at 2, 5, 7.5, 10, and 12.5 K. (b) Magnetic field dependence of critical current density obtained by the data shown in Fig. \ref{f3}(a) assuming a full critical state in the sample.}
\label{f3}
\end{figure}

Figure \ref{f3}(a) depicts the magnetization at several temperatures as a function of field.
From the magnetization hysteresis loop,
we can obtain the critical current density $J_{\textrm{c}}$ using the Bean model with the assumption of field-independent $J_{\textrm{c}}$.
According to the Bean model~\cite{rmp1}, $J_{\textrm{c}}$ is given by
\begin{equation}
	J_c = 20\frac{\Delta M}{a(1-a/3b)}.
	\label{e1}
\end{equation}
where $\Delta M$ is $M_{\textrm{down}} - M_{\textrm{up}}$, $M_{\textrm{up}}$ and $M_{\textrm{down}}$ are the magnetization
when sweeping fields up and down, respectively, $a$ and $b$ are sample widths ($a < b$).
Figure \ref{f3}(b) shows the field dependence of $J_{\textrm{c}}$
obtained from the data shown in Fig. \ref{f3}(a) using Eq. \eqref{e1}
and effective sample dimensions $a \sim$ 320 $\mu$m and $b \sim$ 735 $\mu$m.
A nonmonotonic field dependence of $J_{\textrm{c}}$ with a broad maximum,
fish-tail effect, is observed below 7.5 K as shown in Fig. \ref{f3}(a).
This behavior is similar to Ba(Fe$_{1-x}$Co$_x$)$_2$As$_2$ single crystals~\cite{jpsj,prb6},
YBa$_2$Cu$_3$O$_{7-\delta}$~\cite{prb7}, and La$_{2-x}$Sr$_{x}$CuO$_{4-\delta}$~\cite{phyc}.
The fish-tail effect is more clearly seen in field dependence of critical current densities $J_{\textrm{c}}$ (Fig. \ref{f3}(b)).
Critical current density calculated from $M-H$ curve is estimated to be $1 \times 10^5$ A/cm$^2$ at 5 K under zero field.
Although this value is 1/6 of Ba(Fe$_{1-x}$Co$_{x}$)$_{2}$As$_{2}$ single crystals,
it is still within the range for practical applications~\cite{jpsj}.

\begin{figure}[tb]
\begin{center}
\begin{minipage}{0.49\hsize}
\includegraphics[width=3.5cm,clip]{4a.eps}
\end{minipage}
\vspace{5mm}
\begin{minipage}{0.49\hsize}
\includegraphics[width=3.5cm,clip]{4b.eps}
\end{minipage}
\begin{minipage}{0.49\hsize}
\includegraphics[width=3.5cm,clip]{4c.eps}
\end{minipage}
\vspace{5mm}
\begin{minipage}{0.49\hsize}
\includegraphics[width=3.5cm,clip]{4d.eps}
\end{minipage}
\begin{minipage}{0.49\hsize}
\includegraphics[width=3.8cm,clip]{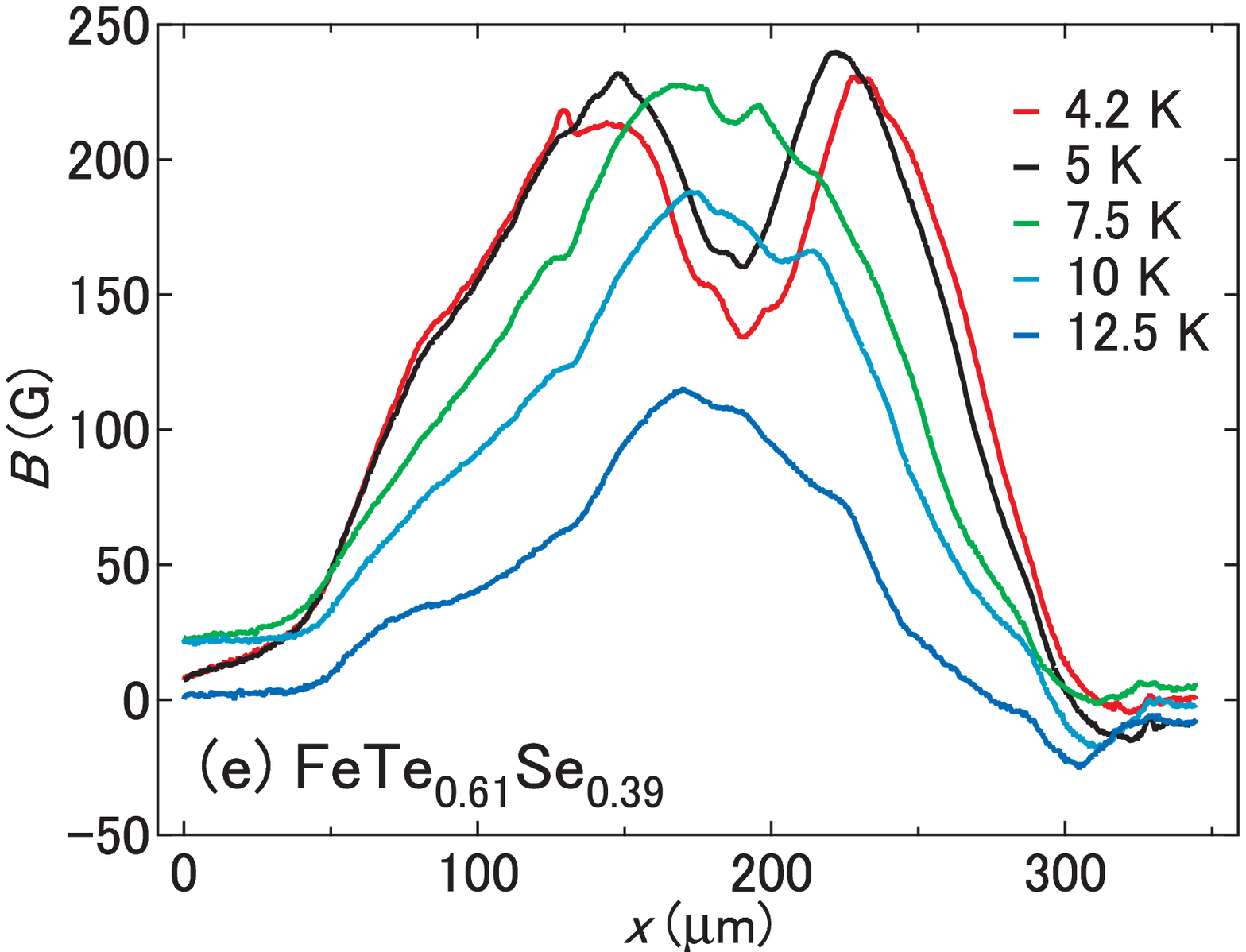}
\end{minipage}
\begin{minipage}{0.49\hsize}
\includegraphics[width=3.5cm,clip]{4f.eps}
\end{minipage}
\end{center}
\caption{(Color online) Magneto-optical images in the remanent state after applying $H$ = 800 Oe in FeTe$_{0.61}$Se$_{0.39}$ at (a) 4.2 K, (b) 7.5 K, (c) 10 K, (d) 12.5 K. (e) Magnetic induction profile along the line drawn in (a). (f) MO image in the remanent state of  an as-grown sample at 10 K. The span of magnetic induction for all the images is from -50 G to 250 G.}
\label{f4}
\end{figure}

The above estimate of $J_{\textrm{c}}$ relies on the assumption that homogeneous current is flowing
within the sample (full critical state).
To examine this assumption,
we made MO imaging of $25$ $\mu$m thick FeTe$_{0.61}$Se$_{0.39}$ in the remanent state at several temperatures.
The state is prepared by applying 800 Oe along $c$ axis for 1 s and removing it after zero-field cooling.
Figures \ref{f4}(a)-(d) show MO images of FeTe$_{0.61}$Se$_{0.39}$ in the remanent state
at 4.2, 7.5, 10, and 12.5 K, respectively.
Figure \ref{f4}(e) shows profiles of the magnetic induction at several temperatures
along the line shown in Fig. \ref{f4}(a).
Since the maximum field of 800 Oe is not enough to create a full critical state at lower temperatures,
the profile is split below 5 K.
At lower temperatures, the MO image shows a typical roof-top pattern indicating a uniform current flow in the sample.
Even at 1.5 K below the bulk $T_{\textrm{c}}$, the uniform current flow is only partly disturbed.
A MO image of an as-grown sample at $T = 10$ K is also shown in Fig. \ref{f4}(f).
It is evident that the sample is superconducting in a limited region and the trapped field is much lower than that in the annealed sample.
All these MO images show that a full critical state is achieved only in the annealed sample,
though there are inhomogeneous penetrations in a limited region.
$J_{\textrm{c}}$ for a thin superconductor is roughly estimated by $J_{\textrm{c}} \sim \Delta B/t$,
where $\Delta B$ is the trapped field and $t$ is the thickness of the sample.
With $\Delta B \sim 230$ G and $t = 25$ $\mu$m,
$J_{\textrm{c}}$ is estimated as $\sim 0.9 \times 10^5$ A/cm$^2$ at 7.5 K,
which is consistent with the value calculated from $M - H$ curve.

At this point, we emphasize the effect of annealing on chemical and superconducting homogeneities in FeTe$_{0.61}$Se$_{0.39}$.
The x-ray diffraction pattern and EDX mapping data reveal that the annealing at low temperature
for an extended period of time is essential to obtain homogeneous samples.
The x-ray diffraction pattern of the annealed sample has much less impurity phases than the as-grown one,
possibly because the low-temperature annealing prohibits phase separation.
The annealing effect on superconductivity is also seen in magnetization.
The temperature dependent magnetization shows a sharp superconducting transition in the annealed sample, though many of as-grown sample do not show superconductivity above 2 K.
This is probably because homogenization as a consequence of annealing makes Te/Se ratio optimum for superconductivity.
MO images demonstrate this improvement more clearly.
The MO image, shown in Fig. \ref{f4}(f) clearly shows large superconducting inhomogeneities in the as-grown crystal.
As seen in Figs. \ref{f4}(a)-(d), annealing makes the sample homogeneous and improves superconductivity.
Besides, we could obtain appreciable hysteresis of magnetization in the annealed crystals.
Such a large hysteresis in $M - H$ curve is exceptional,
since most of previous works reported negligible irreversible magnetization with a large ferro- or ferrimagnetic background~\cite{prb4}.
We also emphasize the simplicity of our method.
We could obtain high quality single crystalline FeTe$_{x}$Se$_{1-x}$ through a simple method of slow cooling followed by annealing,
while Chen \textit{et al.}~\cite{prb3} and  Sales \textit{et al.}~\cite{prb1} reported
FeTe$_{x}$Se$_{1-x}$ crystal growth by Bridgman method.
$\Delta T_{\textrm{c}}$ of our FeTe$_{0.61}$Se$_{0.39}$ is comparable or smaller than the previous report with a similar $T_{\textrm{c}}$ ~\cite{prb1}.
It should be noted, however, that the optimum composition which produces the most homogeneous superconductivity with highest $T_{\textrm{c}}$ in FeTe$_{x}$Se$_{1-x}$
is still unclear,
since the optimum composition is slightly different between our crystal and those reported in Ref. [10].
Further works are necessary to sort out the optimum composition and annealing condition.

In summary, we have prepared high-quality single crystalline FeTe$_{0.61}$Se$_{0.39}$ with $T_{\textrm{c}} \sim 14$ K
by slow-cooling method followed by low-temperature annealing.
Magneto-optical imaging revealed nearly homogeneous current flow in the crystal.
The value of $J_{\textrm{c}}$ is over $1 \times 10^5$ A/cm$^2$ below 5 K under zero field.
$J_{\textrm{c}} - H$ characteristics show a fish-tail effect in the intermediate temperature range,
which is  a promising characteristics for practical superconductors.
The obtained high-quality FeTe$_{x}$Se$_{1-x}$ crystal is expected to serve as a platform
for the investigation of superconducting mechanism in the iron-chalcogenide system.

\begin{acknowledgments}
This work is partly supported by Grant-in-Aid for Scientific Research from Ministry of Education, Culture, Sports, Science and Technology, Japan.
\end{acknowledgments}

\end{document}